\newcommand{\slashedzero}{{0\mkern-9 mu/}}
\newcommand{\hi}{\mathcal O(\hbar)}
\newcommand{\hii}{\mathcal O(\hbar^2)}
\documentclass[%
 reprint,
 amsmath,amssymb,
 aps,
]{revtex4-1}

\usepackage{graphicx}% Include figure files
\usepackage{dcolumn}% Align table columns on decimal point
\usepackage{bm}% bold math
\begin{document}

\preprint{APS/123-QED}

\title{A semi-classical transport equation for the chiral quarks surrounding a SU(2) t'Hooft-Polyakov monopole}% Force line breaks with \\
%\thanks{}%

\author{Feng Li}
\email{fengli@fias.uni-frankfurt.de}
\date{\today}% It is always \today, today,

\begin{abstract}
 A transport equation of the representative quasi-particles is derived, by solving Dirac (or Dyson) equations under the semi-classical approximation, for describing the entangled motions of the red and blue chiral quarks in the vicinity of a SU(2) t'Hooft-Polyakov monopole. The representative quasi-particles have two states characterized by different dispersion relations. A ground state quasi-particle has a zero pole mass, a zero longitudinal inertial mass and a finite transverse inertial mass, while an excited quasi-particle has a finite pole mass and finite inertial masses. At equilibrium, the red and blue currents flow against each other in the transverse direction, and the red and blue axial currents flow against each other in the radial direction.
%\begin{description}
%\end{description}
\end{abstract}

\pacs{Valid PACS appear here}% PACS, the Physics and Astronomy
\maketitle

%\tableofcontents

A t'Hooft-Polyakov monopole is a soliton configuration of a non-abelian gauge field which is first predicted to exist in the system where the gauge field is coupled with an adjoint Higgs field leading to a spontaneous symmetry breaking\cite{Polyakov:1974ek,tHooft:1974kcl,Prasad:1975kr}, and is later predicted to be in a pure gauge theory under an abelian projection\cite{tHooft:1981bkw, Polyakov:1976fu}. Although in the latter case, the location of the monopole depends on the choice of the gauge fixing condition, lattice calculations \cite{DiGiacomo:1999yas,DiGiacomo:1999fb,Carmona:2001ja,Cea:2000zr,Cea:2001jf,Cea:2002wx} suggest that the appearance of the monopole condensation in the QCD vacuum, which is regarded as the reason for quark confinement in the dual Meissner mechanism\cite{Nielsen:1973cs,tHooft:1974kcl,Mandelstam:1974pi,Polyakov:1976fu} where the QCD ground state is analogue to a type-II superconductor, is gauge independent. If that is the case, t'Hooft-Polyakov monopoles should be among the most common objects that a quark will meet. It is therefore interesting to investigate the quark motion in the vicinity of a t'Hooft-Polyakov monopole, which will later be implemented in a partonic transport model where the quark confinement might be hopefully included consistently. In this note, we will derive the Vlasov equation, using a Wigner function (or Dyson-Schwinger) formalism which have been widely employed to derive the partonic transport equations\cite{Elze:1986hq,Elze:1986qd,Elze:1987ii,Vasak:1987um,Elze:1989un,Klevansky:1997wm,Huang:2018wdl,Li:2019rth}, for the chiral quarks moving in the vicinity of a SU(2) t'Hooft-Polyakov monopole.

We start our derivation from a pair of Dirac equations
\begin{eqnarray}
\label{eq:schodinger1}
(i\hbar{\partial\mkern-11 mu/}_{x_1}+{\Omega\mkern-11 mu/}(x_1))G_K(x_1, x_1^\prime) &=& 0, \\
\label{eq:schodinger2}
G_K(x_1, x_1^\prime)(-i\hbar\overleftarrow{\partial\mkern-11 mu/}_{x_1^\prime}+{\Omega\mkern-11 mu/}(x_1^\prime)) &=& 0,
\end{eqnarray}
where $G_K \doteq \langle \Psi(x) \bar\Psi(x^\prime) - \bar\Psi(x^\prime) \Psi(x)\rangle/2$ with $\Psi$ being the quark field is the so-called Keldysh Green's function\cite{Keldysh:1964ud,Rammer:2007zz}, and $\Omega_\mu = \Omega^A_\mu \sigma_A$ with $\sigma$ being the Pauli matrices is the background gluon field. In the following context, we will use the capital alphabets to label the indices of Pauli matrices, the lowercase alphabets to label the indices of $\tau = (\mathbb I, \sigma)$, and the slashed zero $\slashedzero$ to label the zeroth component of $\tau$. The monopole configuration \cite{Polyakov:1974ek,tHooft:1974kcl,Prasad:1975kr,Manton:2004tk} can be written as $\Omega^0_A = 0$, $\Omega^i_A = \epsilon_{Aik}\omega_k$ where $\omega_k = -h(|x|)\hat x^k / |x|$ is a vector pointing to the monopole center. $h(|x|)$ is constructed so that $\Omega_A^\mu$ fulfills the Euler-Lagrangian equation. Notice that by writing the monopole configuration in the above way, we have already chosen a coordinate system where the field lines of $\vec \Omega_A$ are the circles perpendicular to and symmetric about the $A$th coordinate axis, and thus chosen a gauge fixing condition, since the symmetric axes can be rotated under a trivial gauge transformation, i.e., a gauge transformation homotopic to $\mathbb I$. Keep it in mind that all the following discussions are under such a special gauge choice and the obtained transport equation is therefore not gauge invariant.

After doing some math (see Ref.\cite{Klevansky:1997wm, Li:2019rth}), we obtain from Eq.(\ref{eq:schodinger1}) and Eq.(\ref{eq:schodinger2}) that 
\begin{eqnarray}
\label{eq:DysonApprox1}
0&=&\frac i 2\hbar\partial_{X^\mu}\{\gamma^\mu,{\widetilde G}(X,p)\}+[{k\mkern-11 mu/},{\widetilde G}(X,p)]\nonumber\\
&&-\frac i 2\hbar\{\partial_{X^\mu}{\Omega\mkern-11 mu/}(X),\partial_{p_\mu}{\widetilde G}(X,p)\}+\mathcal O(\hbar^2),\\
\label{eq:DysonApprox2}
0&=&\frac i 2\hbar\partial_{X^\mu}[\gamma^\mu,{\widetilde G}(X,p)]+\{{k\mkern-11 mu/},{\widetilde G}(X,p)\}\nonumber\\
&&-\frac i 2\hbar[\partial_{X^\mu}{\Omega\mkern-11 mu/}(X),\partial_{p_\mu}{\widetilde G}(X,p)]+\mathcal O(\hbar^2),
\end{eqnarray}
where $k_\slashedzero = p$, $k_A = \Omega_A$ and
\begin{eqnarray*}
\widetilde G(X,p) \doteq \int d^4x e^{ipx/\hbar}G_K(X+x/2, X-x/2), \\
\end{eqnarray*}
which can be decomposed in both the spinor and color spaces as
\begin{equation}
\label{eq:GComponent}
    \widetilde G = S_a \tau^a + iP_a \tau^a\gamma_5 + V_\mu^a \tau_a\gamma^\mu + A^a_\mu \tau_a\gamma^\mu\gamma_5 + \frac 1 2 J^a_{\mu\nu}\tau_a\sigma^{\mu\nu}
\end{equation}
with $\sigma^{\mu\nu}$ being $\frac i 2 [\gamma^\mu, \gamma^\nu]$. The components in Eq.(\ref{eq:GComponent}) have their physical meanings, i.e.,
\begin{eqnarray}
   j_V^\mu (X) &=& {\rm tr} \gamma^\mu G_K(X,X) = \int d^4p V^\mu_\slashedzero,\\
    j_A^\mu (X) &=& {\rm tr} \gamma_5 \gamma^\mu G_K(X,X) = \int d^4p A^\mu_\slashedzero,\\
    q_V^\mu (X) &=& {\rm tr} \gamma^\mu \tau_3 G_K(X,X) = \int d^4p V^\mu_3,\\
    q_A^\mu (X) &=& {\rm tr} \gamma_5 \gamma^\mu \tau_3 G_K(X,X) = \int d^4p A^\mu_3,
\end{eqnarray}
where $j_V$, $j_A$, $q_V$ and $q_A$ are the color averaged current density, the color averaged axial current density, the difference between the current densities of the red and blue quarks, and the difference between the axial current densities of the red and blue quarks, respectively.
\begin{widetext}
Using the identities
\begin{eqnarray*}
\{U\gamma^\mu,\widetilde G\} &=& \{U,V^\mu\}+[U,A^\mu]\gamma_5 + (g^{\mu\nu}\{U,S\} + i[U,J^{\mu\nu}])\gamma_\nu +\left(ig^{\mu\nu}[U,P] - \frac 1 2 \epsilon^{\mu\nu\rho\sigma}\{U,J_{\rho\sigma}\}\right)\gamma_\nu \gamma_5 \nonumber\\
&& + \left(-ig^{\mu\rho}[U,V^\sigma] + \frac 1 2 \epsilon^{\mu\nu\rho\sigma}\{U,A_\nu\}\right)\sigma_{\rho\sigma},\\{}
 [U\gamma^\mu,\widetilde G] &=& [U,V^\mu]+\{U,A^\mu\}\gamma_5  + (g^{\mu\nu}[U,S] + i\{U,J^{\mu\nu}\})\gamma_\nu + \left(ig^{\mu\nu}\{U,P\} - \frac 1 2 \epsilon^{\mu\nu\rho\sigma}[U,J_{\rho\sigma}]\right)\gamma_\nu \gamma_5 \nonumber\\
&&+ \left(-ig^{\mu\rho}\{U,V^\sigma\} +\frac 1 2 \epsilon^{\mu\nu\rho\sigma}[U,A_\nu]\right)\sigma_{\rho\sigma},
\end{eqnarray*}
where $U$ is a $2\times 2$ matrix defined in the color space, we decompose Eq. (\ref{eq:DysonApprox1}) and Eq. (\ref{eq:DysonApprox2}) in the spinor space, and write down only the equations containing $V$ and $A$, which are the scalar, pseudo-scalar, and tensor components of Eq.(\ref{eq:DysonApprox1}) and Eq.(\ref{eq:DysonApprox2}), i.e.,
\begin{eqnarray}
\label{eq:scalar0}
0&=&2\hbar\partial_{X^\mu}V_a^\mu \tau^a - \hbar\partial_{X^\mu}\Omega^a_\nu \partial_{p_\mu}V_b^\nu\{\tau^a,\tau^b\} - 2ik^a_\mu V^\mu_b [\tau^a,\tau^b]+\hii,\\
0&=&2k_\mu^a A_b^\mu \{\tau^a,\tau^b\} - i \hbar \partial_{X^\mu}\Omega_\nu^a \partial_{p_\mu}A_b^\nu[\tau^a,\tau^b] + \hii,\\
0&=&\hbar\epsilon^{\mu\nu\rho\sigma}\partial_{X^\mu}A_\nu^a\tau^a-\left(k^\rho_a V^\sigma_b-k^\sigma_a V^\rho_b + \frac \hbar 2 \epsilon^{\mu\nu\rho\sigma}\partial_{X^\tau}\Omega^a_\mu \partial_{p_\tau} A_\nu^b \right)\{\tau^a,\tau^b\} \nonumber\\
&&-i\left(\epsilon^{\mu\nu\rho\sigma}k^a_\mu A^b_\nu-\frac \hbar 2 \partial_{X^\tau}\Omega^\rho_a \partial_{p_\tau}V_b^\sigma + \frac \hbar 2 \partial_{X^\tau}\Omega^\sigma_a \partial_{p_\tau}V_b^\rho\right)[\tau^a,\tau^b] + \hii,\\
0&=&2\hbar\partial_{X^\mu}A_a^\mu \tau^a - \hbar\partial_{X^\mu}\Omega^a_\nu \partial_{p_\mu}A_b^\nu\{\tau^a,\tau^b\} - 2ik^a_\mu A^\mu_b [\tau^a,\tau^b] +\hii,\\
0&=&2k_\mu^a V_b^\mu \{\tau^a,\tau^b\} - i \hbar\partial_{X^\mu}\Omega_\nu^a \partial_{p_\mu}V_b^\nu[\tau^a,\tau^b]+\hii,\\
\label{eq:tensor1}
0&=&\hbar\epsilon^{\mu\nu\rho\sigma}\partial_{X^\mu}V_\nu^a\tau^a-\left(k^\rho_a A^\sigma_b-k^\sigma_a A^\rho_b + \frac \hbar 2 \epsilon^{\mu\nu\rho\sigma}\partial_{X^\tau}\Omega^a_\mu \partial_{p_\tau} V_\nu^b \right)\{\tau^a,\tau^b\}\nonumber\\
&&-i\left(\epsilon^{\mu\nu\rho\sigma}k^a_\mu V^b_\nu-\frac \hbar 2 \partial_{X^\tau}\Omega^\rho_a \partial_{p_\tau}A_b^\sigma + \frac \hbar 2 \partial_{X^\tau}\Omega^\sigma_a \partial_{p_\tau}A_b^\rho\right)[\tau^a,\tau^b] +\hii.
\end{eqnarray}
All the other equations are about $S$, $P$ and $J$, which are not concerned in the scope of this work. Eq.(\ref{eq:scalar0}-\ref{eq:tensor1}) can be simplified into two groups of decoupled equations by substituting $V$ and $A$ with $\mathcal F_+ + \mathcal F_-$ and $\mathcal F_+ - \mathcal F_-$ respectively, and they are
\begin{eqnarray}
\label{eq:F1}
0&=&2\hbar\partial_{X^\mu}\mathcal F_{\pm a}^\mu \tau^a - \hbar \partial_{X^\mu}\Omega^a_\nu \partial_{p_\mu}\mathcal F_{\pm b}^\nu\{\tau^a,\tau^b\} - 2ik^a_\mu \mathcal F^\mu_{\pm b} [\tau^a,\tau^b]+\hii,\\
0&=&2k_\mu^a \mathcal F_{\pm b}^\mu \{\tau^a,\tau^b\} - i \hbar \partial_{X^\mu}\Omega_\nu^a \partial_{p_\mu}\mathcal F_{\pm b}^\nu[\tau^a,\tau^b] +\hii,\\
\label{eq:F3}
0&=&\pm\hbar\epsilon^{\mu\nu\rho\sigma}\partial_{X^\mu}\mathcal F_{\pm\nu}^a\tau^a-\left(k^\rho_a \mathcal F^\sigma_{\pm b}-k^\sigma_a \mathcal F^\rho_{\pm b} \pm \frac \hbar 2 \epsilon^{\mu\nu\rho\sigma}\partial_{X^\tau}\Omega^a_\mu \partial_{p_\tau} \mathcal F_{\pm\nu}^b \right)\{\tau^a,\tau^b\}\nonumber\\
&&-i\left(\pm\epsilon^{\mu\nu\rho\sigma}k^a_\mu \mathcal F^b_{\pm\nu}-\frac \hbar 2 \partial_{X^\tau}\Omega^\rho_a \partial_{p_\tau}\mathcal F_{\pm b}^\sigma + \frac \hbar 2 \partial_{X^\tau}\Omega^\sigma_a \partial_{p_\tau}\mathcal F_{\pm b}^\rho\right)[\tau^a,\tau^b] +\hii,
\end{eqnarray}
where the dependence of $\mathcal F_\pm$ on $(X,p)$ characterizes the motions of the left and right handed quarks, respectively. We further decompose Eq.(\ref{eq:F1}-\ref{eq:F3}) in the color space and obtain
\begin{eqnarray}
\label{eq:S0}
0 &=& \partial_{X^\mu}\mathcal F^\mu_{\pm\slashedzero}-\partial_{X^\mu}\Omega^A_\nu \partial_{p_\mu} \mathcal F_{\pm A}^\nu +\hii, \\
0 &=& \hbar\partial_{X^\mu}\mathcal F^\mu_{\pm A}-\hbar\partial_{X^\mu}\Omega^A_\nu \partial_{p_\mu} \mathcal F_{\pm \slashedzero}^\nu + 2\epsilon_{ABC} \Omega^B_\mu \mathcal F_{\pm C}^\mu+\hii,\\
0 &=& p_\mu \mathcal F_{\pm\slashedzero}^\mu + \Omega^A_\mu \mathcal F_{\pm A}^\mu + \hii,\\
0 &=& p_\mu \mathcal F_{\pm A}^\mu + \Omega^A_\mu \mathcal F_{\pm \slashedzero}^\mu + \frac \hbar 2 \epsilon_{ABC} \partial_{X^\mu}\Omega_\nu^B \partial_{p_\mu} \mathcal F^\nu_{\pm C} + \hii, \\
0 &=& \pm \hbar \epsilon^{\mu\nu\rho\sigma}\left(\partial_{X^\mu}\mathcal F_{\pm\nu}^\slashedzero - \partial_{X^\tau}\Omega_\mu^A \partial_{p_\tau}\mathcal F_{\pm \nu}^A\right) - 2(k^\rho_a \mathcal F_{\pm a}^\sigma - k^\sigma_a \mathcal F^\rho_{\pm a}) + \hii,\\
\label{eq:TA}
0 &=& \pm \epsilon^{\mu\nu\rho\sigma}\left(\hbar\partial_{X^\mu}\mathcal F_{\pm\nu}^A - \hbar\partial_{X^\tau}\Omega_\mu^A \partial_{p_\tau}\mathcal F_{\pm \nu}^\slashedzero + 2 \epsilon_{ABC}\Omega_\mu^B \mathcal F_{\pm \nu}^C\right) - 2(p^\rho \mathcal F_{\pm A}^\sigma - p^\sigma \mathcal F^\rho_{\pm A} + \Omega_A^\rho\mathcal F_{\pm \slashedzero}^\sigma - \Omega_A^\sigma \mathcal F_{\pm\slashedzero}^\rho)\nonumber\\
&&- \hbar\epsilon_{ABC}\left(\partial_{X^\tau} \Omega_B^\rho \partial_{p_\tau}\mathcal F^\sigma_{\pm C} - \partial_{X^\tau} \Omega_B^\sigma \partial_{p_\tau}\mathcal F^\rho_{\pm C} \right) + \hii.
\end{eqnarray}
\end{widetext}

\begin{figure}
    %\centering
    \includegraphics[width=0.45\textwidth]{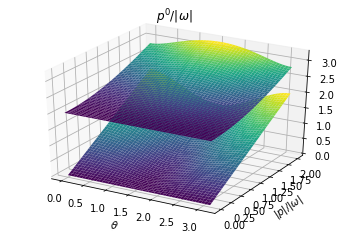}
    \caption{Dispersion relation of the representative quasi-particle from whose distribution function we can calculate both the current and axial current densities of both the red and blue quarks. $\theta$  is the angle between momentum $\vec p$ and $\vec\omega$.}
    \label{fig:dispersion}
\end{figure}
It is explained in Ref.\cite{Li:2019rth} why the higher order term in Eq.(\ref{eq:S0}) is $\hii$ rather than $\hi$. To find a semi-classical solution to Eq.(\ref{eq:S0}-\ref{eq:TA}), we expand $\mathcal F_{\pm a}^\mu$ as $\mathcal F_{\pm a}^\mu = \mathcal F_{\pm a}^{(0)\mu} + \hi$, neglect all the terms proportional to $\hbar$ in Eq.(\ref{eq:S0}-\ref{eq:TA}) and obtain
\begin{eqnarray}
\label{eq:0S0}
0 &=& \partial_{X^\mu}\mathcal F^{(0)\mu}_{\pm\slashedzero}-\partial_{X^\mu}\Omega^A_\nu \partial_{p_\mu} \mathcal F_{\pm A}^{(0)\nu} +\hi, \\
\label{eq:0SA}
0 &=& \epsilon_{ABC} \Omega^B_\mu \mathcal F_{\pm C}^{(0)\mu}+\hi,\\
\label{eq:0A0}
0 &=& p_\mu \mathcal F_{\pm\slashedzero}^{(0)\mu} + \Omega^A_\mu \mathcal F_{\pm A}^{(0)\mu} + \hi,\\
\label{eq:0AA}
0 &=& p_\mu \mathcal F_{\pm A}^{(0)\mu} + \Omega^A_\mu \mathcal F_{\pm \slashedzero}^{(0)\mu} + \hi, \\
\label{eq:0T0}
0 &=& k^\rho_a \mathcal F_{\pm a}^{(0)\sigma} - k^\sigma_a \mathcal F^{(0)\rho}_{\pm a} + \hi,\\
\label{eq:0TA}
0 &=& p^\rho \mathcal F_{\pm A}^{(0)\sigma} - p^\sigma \mathcal F^{(0)\rho}_{\pm A} + \Omega_A^\rho\mathcal F_{\pm \slashedzero}^{(0)\sigma} \nonumber\\
&& - \Omega_A^\sigma \mathcal F_{\pm\slashedzero}^{(0)\rho} \mp \epsilon^{\mu\nu\rho\sigma} \epsilon_{ABC}\Omega_\mu^B \mathcal F_{\pm \nu}^{(0)C} +\hi.
\end{eqnarray}
Eq.(\ref{eq:0SA}-\ref{eq:0TA}) are the linear equations of $\mathcal F_{\pm a}^{(0)\mu}$ which both have 16 components. We first solve Eq.(\ref{eq:0T0}-\ref{eq:0TA}) which are of rank 14 if $\Omega_\mu$ is abelian, i.e., $\Omega_\mu = \Omega^3_\mu \sigma_3$. It would allow us to express all the components of $\mathcal F_{\pm a}^{(0)\mu}$ as a superposition of two functions which can be regarded as the distribution functions of the red and blue quarks respectively. Therefore, if the background field is abelian, the motions of the red and blue quarks can be described by a pair of decoupled transport equations. However, if we substitute $\Omega_\mu^A$ with the t'Hooft-Polyakov monopole configuration, i.e., $\Omega^0_A = 0$ and $\Omega^i_A = \epsilon_{Aij}\omega_j$, the rank of Eq.(\ref{eq:0T0}-\ref{eq:0TA}) is 15, which means the components of $\mathcal F_{\pm a}^\mu$ can be expressed using only one function, and the motions of the red and blue quarks are therefore entangled, which is a feature of confinement. The solution, obtained with the help of some symbolic calculation software such as sympy, to Eq.(\ref{eq:0T0}-\ref{eq:0TA}) for $\Omega_A^\mu$ being the monopole configuration is
\begin{eqnarray}
\label{eq:F0}
\mathcal F_{\pm\slashedzero}^{(0)\mu} &=& \left[(p^2+\Omega_A^\nu \Omega^A_\nu)p^\mu-2p^\nu\Omega_\nu^A\Omega^\mu_A\right]F_\pm,\\
\label{eq:FA}
\mathcal F_{\pm A}^{(0)\mu} &=& \left[p^2\Omega_A^\mu-2\Omega^\nu_A p_\nu p^\mu \mp \epsilon_{ABC}\epsilon^{\mu\nu\rho\sigma} \Omega^B_\nu \Omega^C_\rho p_\sigma \right]F_\pm, \nonumber\\
\end{eqnarray}
where $F_\pm$ contain the distribution functions of the left and right handed quasi-particles respectively, from which we can calculate both the current and axial current densities of both the blue and red quarks. We therefore call it the representative quasi-particle. Eq.(\ref{eq:0SA}) and Eq.(\ref{eq:0AA}) are automatically fulfilled by Eq.(\ref{eq:F0}-\ref{eq:FA}). The dispersion relation of the representative quasi-particle, obtained by substituting $\mathcal F$ in Eq.(\ref{eq:0A0}) with the right hand side of Eq.(\ref{eq:F0}) and Eq.(\ref{eq:FA}), is
\begin{equation}
    (p^2-2\vec\omega^2)^2 = 4[\vec \omega^4 + \vec p^2 \vec\omega^2 - (\vec p \cdot \vec\omega)^2],
\end{equation}
or
\begin{equation}
   p^0 = \sqrt{\vec p^2 + 2\vec \omega^2 \pm 2\sqrt{\vec\omega^4 + \vec p^2 \vec\omega^2 - (\vec p \cdot \vec\omega)^2}},
\end{equation}
and is plotted in Fig.(\ref{fig:dispersion}) where $\theta$ is the angle between $\vec p$ and $\vec \omega$. It shows that the representative quasi-particle can be at two energy-levels. For $|\vec p| \ll |\vec \omega|$,
\begin{equation}
\label{eq:energy_smallp}
    p^0 \approx \sqrt{p_\perp^2 + p_\parallel^2 + 2 \vec\omega^2 \pm (2\vec\omega^2 + p_\perp^2 - p_\perp^4 / 4\vec\omega^2)}
\end{equation}
where $p_\parallel$ and $p_\perp$ are the $\vec p$ components parallel and perpendicular to $\vec\omega$ respectively. The masses of the representative quasi-particle can be easily read-off from Eq.(\ref{eq:energy_smallp}). For the quasi-particle at the upper level, its pole-mass, defined as $\lim_{|\vec p| \to 0} p^0$, is $2|\vec \omega|$, its longitudinal inertial mass, defined as $\lim_{p_\parallel \to 0} \lim_{p_\perp \to 0} (\partial^2 p^0 / \partial p_\parallel^2)^{-1}$, is $2|\vec\omega|$, and its transverse inertial mass, defined as $\lim_{p_\perp \to 0} \lim_{p_\parallel \to 0}(\partial^2 p^0 / \partial p_\perp^2)^{-1}$, is $|\vec\omega|$. Please notice the order of taking limitations. The longitudinal inertial mass characterizes the slow motion in the radial direction, therefore $p_\perp$ is first taken zero so that $\vec p$ is in the radial direction, vice versa. For the quasi-particle at the ground level, both its pole and longitudinal inertial masses are zero, and its transverse inertial mass is $|\vec \omega|$. A representative quasi-particle might be excited from the ground level to the upper level due to collisions, and therefore gains finite pole and longitudinal inertial masses. It might be the mechanism how a chiral is weighted. 

Adding the dispersion relation in Eq.(\ref{eq:F0}) and Eq.(\ref{eq:FA}), we obtain
\begin{widetext}
\begin{eqnarray}
\label{eq:f0}
\mathcal F_{\pm\slashedzero}^{(0)\mu} &=& \delta((p^2-2\vec\omega^2)^2 - 4[\vec \omega^4 + \vec p^2 \vec\omega^2 - (\vec p \cdot \vec\omega)^2])\left[(p^2+\Omega_A^\nu \Omega^A_\nu)p^\mu-2p^\nu\Omega_\nu^A\Omega^\mu_A\right]f_\pm,\\
\label{eq:fA}
\mathcal F_{\pm A}^{(0)\mu} &=& \delta((p^2-2\vec\omega^2)^2 - 4[\vec \omega^4 + \vec p^2 \vec\omega^2 - (\vec p \cdot \vec\omega)^2]) \left[p^2\Omega_A^\mu-2\Omega^\nu_A p_\nu p^\mu \mp \epsilon_{ABC}\epsilon^{\mu\nu\rho\sigma} \Omega^B_\nu \Omega^C_\rho p_\sigma \right]f_\pm,
\end{eqnarray}
where $f_\pm$ is the distribution function of the left and right handed representative quasi-particle respectively, which fulfills the transport equation
\begin{eqnarray}
\label{eq:transport}
0&=&\left\{\left[(p^2+\Omega_A^\nu \Omega^A_\nu) p^\mu - 2p^\nu\Omega_\nu^A \Omega_A^\mu\right]\partial_{X^\mu}f_\pm - \partial_{X^\mu}\Omega_\nu^A \left[p^2\Omega_A^\nu-2\Omega_A^\tau p_\tau p^\nu\right]\partial_{p_\mu} f_\pm \right\}\nonumber\\
&&\times \delta((p^2-2\vec\omega^2)^2 - 4[\vec \omega^4 + \vec p^2 \vec\omega^2 - (\vec p \cdot \vec\omega)^2]),
\end{eqnarray}
\end{widetext}
obtained by substituting $\mathcal F$ in Eq.(\ref{eq:0S0}) with the right hand side of Eq.(\ref{eq:f0}, \ref{eq:fA}).

A reasonable assumption is that, at equilibrium, $f_+^{(0)} = f_-^{(0)}$ are both even functions of both $|\vec p|$ and $\vec p \cdot \vec \omega$, and the difference between the axial 3-current densities of the red and blue quarks
\begin{eqnarray}
q_A^{(0)i} (X) &=& \int d^4p \left(\mathcal F^{(0)\mu}_{+3} - \mathcal F^{(0)\mu}_{-3}\right)\nonumber\\
&=& \int \frac{d^3\vec p}{4p_0^3}2p^0\omega^i\omega_3(f^{(0)}_+ + f^{(0)}_-),
\end{eqnarray}
is in the radial direction, while the difference between the 3-current densities of the red and blue quarks
\begin{eqnarray}
q_V^{(0)i} (X) &=& \int d^4p \left(\mathcal F^{(0)\mu}_{+3} + \mathcal F^{(0)\mu}_{-3}\right)\nonumber\\
&=& \int \frac{d^3\vec p}{4p_0^3}p^2\epsilon_{3ij}\omega_j(f^{(0)}_+ + f^{(0)}_-)
\end{eqnarray}
is in the transverse direction. Besides, the color averaged 3-current and axial 3-current densities are both zero at equilibrium. So the equilibrium scenario is following, in the radial direction, the red left-handed quarks are moving against the red right-handed quarks but with the blue right-handed quarks, while in the transverse direction, the red left-handed quarks are moving with the red right-handed quarks but against the blue right-handed quarks.

In the non-equilibrium case, Eq.(\ref{eq:transport}) can be solved using the test particle method\cite{Wong:1982zzb} where $f_\pm(\vec x, \vec p, t)$ is replaced with $N^{-1}\sum_a \delta(\vec x - \vec x_a(t)) \delta(\vec p - \vec p_a(t) )$ with $\vec x_a$ and $\vec p_a$ being the positions and the momenta of the test particles which fulfill the equations of motion:
\begin{eqnarray}
(p^2_a-2\vec\omega^2) p^0_a \dot{\vec x}_a &=& p^2_a \vec p_a - 2 (\vec\omega\cdot\vec p_a)\vec\omega, \\
(p^2_a-2\vec\omega^2) p^0_a \dot{\vec p}_a &=& -p_a^{02}\vec\nabla \vec\omega^2 + 2(\vec p_a \cdot \vec\omega)\vec\nabla (\vec p_a \cdot \vec\omega),
\end{eqnarray}
derived from Eq.(\ref{eq:transport}).

In summary, we solve Dirac (or Dyson) equations for the chiral quarks moving in the vicinity of a SU(2) t'Hooft-Polyakov monopole under the semi-classical approximation and find that the motions of the red and blue quarks are entangled, and can be described by the evolution of the distribution function of representative quasi-particles which have two states characterized by different dispersion relations. A ground state quasi-particle has a zero pole mass, a zero longitudinal inertial mass and a finite transverse inertial mass, while a quasi-particle at the excited state has a finite pole mass and finite inertial masses. At equilibrium, the red and blue currents flow against each other in the transverse direction, and the red and blue axial currents flow against each other in the radial direction. We will extend our derivation to the order of $\hbar$ in the near future to study the quantum effect on the motions of the chiral quarks in the vicinity of the monopole. 

\bibliography{references}% Produces the bibliography via BibTeX.

\end{document}